\documentclass[12pt,a4paper]{article}
\usepackage[latin1]{inputenc}
\usepackage[table]{xcolor}

\usepackage{booktabs}

\usepackage{graphicx}

\usepackage{amsfonts}
\usepackage{amsmath}
\usepackage{amssymb}
\usepackage{theorem}
\usepackage{upgreek}

\theoremstyle{change}
\theorembodyfont{\slshape}

\theorembodyfont{\upshape\small}

\theorembodyfont{\ttfamily}

\newcommand{\ba}{\begin{equation}}
\newcommand{\ea}{\end{equation}}

\usepackage{dsfont}

\newcommand{\bin}{\textup{Bin}}

\newcommand{\poi}{\textup{Poi}}
\newcommand{\nb}{\textup{NB}}

\newcommand{\bbn}{\mathbb{N}}
\newcommand{\bbr}{\mathbb{R}}

\newcommand{\iid}{i.\,i.\,d.}
\newcommand{\ie}{i.\,e., }
\newcommand{\eg}{e.\,g., }

\usepackage{natbib}

\usepackage{hyperref}

\begin{document}



\parindent 0cm

\title{Stein EWMA Control Charts for Count Processes}
\author{
Christian H.\ Wei\ss\thanks{
Helmut Schmidt University, Department of Mathematics and Statistics, Hamburg, Germany.}\ \thanks{Corresponding author. E-Mail: \href{mailto:weissc@hsu-hh.de}{\nolinkurl{weissc@hsu-hh.de}}. ORCID: \href{https://orcid.org/0000-0001-8739-6631}{0000-0001-8739-6631}}
}

\maketitle

\begin{abstract}
\noindent
The monitoring of serially independent or autocorrelated count processes is considered, having a Poisson or (negative) binomial marginal distribution under in-control conditions. Utilizing the corresponding Stein identities, exponentially weighted moving-average (EWMA) control charts are constructed, which can be flexibly adapted to uncover zero inflation, over- or underdispersion. The proposed Stein EWMA charts' performance is investigated by simulations, and their usefulness is demonstrated by a real-world data example from health surveillance. 

\medskip
\noindent
\textsc{Key words:}
attributes data; average run lengths; count time series; EWMA control charts; Stein identity
\end{abstract}

\section{Introduction}
\label{Introduction}
The sequential monitoring of count processes $(X_t)_{t\in\bbn=\{1,2,\ldots\}}$ (\ie where the~$X_t$ have a quantitative range contained in $\bbn_0=\{0,1,\ldots\}$) is of utmost importance in many application areas, such as the quality control of manufactured items (counts of defects or non-conformities) or health surveillance (counts of infections or hospital admissions); see \citet{mont09,weiss15,weiss18} for examples and references. The primary tool for managing statistical process control (SPC) for $(X_t)$ is the use of attributes control charts. These charts involve plotting specific statistics sequentially until signals indicate the necessity for corrective action. These statistics are computed in an online manner from the sequence of incoming counts~$(X_t)$. The control chart triggers a signal (``alarm'') if the plotted statistic violates the specified control limits (CLs). The CLs are derived from the so-called in-control model of~$(X_t)$, \ie a stochastic model that assumes~$(X_t)$ to operate under stable conditions. Hence, an alarm (a violation of the CLs) is interpreted as an indication of a possible process deterioration (out-of-control situation). Therefore, if the alarm occurs when the process $(X_t)$ is genuinely out of control, it is considered a true alarm. On the other hand, an alarm under in-control conditions is regarded as a false alarm. It stands to reason that CLs should be chosen to avoid false alarms for as long as possible, and to receive a true alarm as soon as possible. The default metric for evaluating these durations until alarm is the average run length (ARL), \ie the expected number of plotted statistics until the first alarm. For a more detailed description of the aforementioned terms and concepts as well as for further references, the reader may consult SPC textbooks such as \citet{mont09,qiu14}.

\smallskip
Various control charts for diverse count processes have been developed during the last decades, covering serially independent or autocorrelated counts, and counts having full~$\bbn_0$ as their range (unbounded counts) or just the finite subset $\{0,\ldots,n\}$ with specified $n\in\bbn$ (bounded counts). 
The most simple type of control chart is obtained by plotting the counts~$X_t$ themselves against appropriately chosen CLs, which is called a c-chart if monitoring unbounded counts and an np-chart for bounded counts. Such types of Shewhart control chart can be classified as being memory-less as the $t$th plotted statistic does not comprise information about earlier counts (at least not beyond the mere effect of autocorrelation). Consequently, while Shewhart charts might be quick in detecting a sudden strong process shift, they are slow in detecting small process changes. Therefore, also several memory-type control charts for counts have been proposed, where in the present research, our focus is on exponentially weighted moving-average (EWMA) charts. Some recent references on EWMA-type control charts for counts are \citet{gan90,borror98,weiss11,rakitzis15,morais18,morais20,anastasopoulou22a,anastasopoulou22b}. Further references (also on other types of count control charts) can be found in \citet{weiss15,weiss18,alevizakos20}. The default version of the EWMA chart plots the statistics
\ba
\label{ewmarecursion}
Z_0=\mu_0,\quad
Z_t\ =\ \lambda\cdot X_t\ +\ (1-\lambda)\cdot Z_{t-1}\quad\text{for } t=1,2,\ldots
\ea
against a specified lower and upper CL (LCL and UCL, respectively), where $\mu_0>0$ denotes the in-control mean of $(X_t)$. The smoothing parameter $\lambda\in (0,1]$ in \eqref{ewmarecursion} controls the strength of the memory (the smaller~$\lambda$, the stronger the memory). If $\lambda=1$, then \eqref{ewmarecursion} reduces to the c- or np-chart, respectively.

\medskip
All the aforementioned count EWMA charts are mainly designed to detect shifts in the process mean, although they may sometimes (``accidentally'') also react to increases in variance or changes in the autocorrelation structure. In the present research, however, our focus is on ``more sophisticated'' out-of-control scenarios, namely where the mean changes together with further distributional properties (\ie where the process change cannot be traced back to a sole change in the mean parameter), or where ``purely distributional changes'' (not affecting the mean) happen. Such ``distributional changes'' might be increases or decreases in dispersion compared to the in-control model (overdispersion or underdispersion, respectively), or an excessive number of zero counts (zero inflation), to mention those being most relevant in practice. The basic idea of our approach is as follows. Many common count distributions can be characterized by a type of moment identity, referred to as \emph{Stein identity}, which has to hold for a large class of functions (\eg all bounded functions on~$\bbn_0$). The idea to develop such identities dates back to \citet{stein72,stein86}, and further contributions and references can be found in \citet{sudheesh09,sudheesh12,landsman16}. Recently, such Stein identities were successfully used to develop powerful goodness-of-fit (GoF) tests for counts, see \citet{betsch22,weissetal23} for Stein-type GoF-tests for independent and identically distributed (\iid) counts, and \citet{aleksandrov22,aleksandrov22b} for tests for count time series. Thus, it suggests itself to utilize these Stein identities also for developing sequential test procedures, namely count control charts for relevant types of in-control model. In a first research \citep[see][]{weiss23}, this idea was tried for the special case of \iid\ Poisson counts, and the achieved ARL performance was quite appealing. This motivates to develop and investigate Stein-based control charts for count data on a much broader scale, namely for various different count distributions and not only for \iid\ but also for time series data. More precisely, we focus on the three most common count distributions in practice, namely Poisson (Poi), negative binomial (NB), and binomial (Bin), and we include first-order autoregressive (AR$(1)$) models having either Poi-, NB-, or Bin-distributed marginal distributions in our research. 

\smallskip
The outline of this chapter is as follows. In Section~\ref{Count Models and Stein Identities}, we briefly present the count models used for this research, and we provide the Stein identities for the Poi-, NB-, and Bin-distribution. These are used in Section~\ref{Stein EWMA Charts for Counts} to construct novel EWMA-type control charts for counts, where the chart design with respect to diverse out-of-control scenarios is discussed in detail. 
In Section~\ref{Simulation-based Performance Analyses}, results from a simulation study are presented, which allow to analyze the Stein EWMA charts' ARL performance. 
Section~\ref{An Illustrative Data Example} then investigates a real-world data example on registrations in the emergency department of a children's hospital, which illustrates the application and interpretation of the novel Stein EWMA charts in practice. 
Finally, Section~\ref{Conclusions} concludes the article and discusses possible directions for future research.

\section{Count Models and Stein Identities}
\label{Count Models and Stein Identities}
The two most common distributions for unbounded counts are the Poi- and NB-distribution, while the Bin-distribution is the default choice for bounded counts; see \citet{johnson05} for details and properties. The $\poi(\mu)$-distribution with mean~$\mu>0$ is known to be equidispersed, \ie its variance~$\sigma^2$ satisfies $\sigma^2=\mu$. By contrast, $\nb(\nu, \frac{\nu}{\nu+\mu})$ with $\nu,\mu>0$ exhibits overdispersion relative to the Poi-distribution, because its variance $\sigma^2=(1+\frac{\mu}{\nu})\, \mu$ always exceeds the mean. The $\bin(n,\mu/n)$-distribution with $n\in\bbn$ and $\mu\in (0,n)$, in turn, refers to counts having the bounded range $\{0,\ldots,n\}$. 

\smallskip
There is a huge variety of time series models related to the Poi-, NB-, or Bin-distribution, see \citet{weiss18} for a comprehensive overview. Here, many models use a \emph{conditional} Poi-, NB-, or Bin-distribution (regression models) while the marginal distribution does not belong to any parametric model family. However, to be able to apply a Stein identity to given data, we need to specify the corresponding \emph{marginal} distribution. Regarding the count time series models proposed so far \cite{weiss18}, only a considerably smaller number of models has a Poi-, NB-, or Bin-\emph{marginal} distribution. For the sake of studying possible effects of serial dependence on the charts' performance, we shall consider three of the AR$(1)$-type processes surveyed by \citet{weiss08}, whose model definitions use so-called thinning operators as integer substitutes of the multiplication:
\begin{itemize}
	\item the Poi-INAR$(1)$ process (integer AR) defined by
\ba
\label{recINAR1}
X_t\ =\ \rho\circ X_{t-1} + \epsilon_t
\quad\text{with \iid\ }
\epsilon_t\sim\poi\big(\mu(1-\rho)\big),
\ea
	\item the NB-IINAR$(1)$ process (iterated-thinning INAR) defined by
\ba
\label{recIINAR1}
X_t\ =\ \rho\circledast_\pi X_{t-1} + \epsilon_t
\quad\text{with \iid\ }
\epsilon_t\sim\nb(\nu,\pi)
\text{ and } \pi=\tfrac{\nu}{\mu(1-\rho)+\nu},
\ea
	\item the BinAR$(1)$ process defined by
\ba
\label{recBinAR1}
X_t\ =\ \alpha\circ X_{t-1} + \beta\circ (n-X_{t-1})
\quad\text{with }
\beta=(1-\rho)\, \tfrac{\mu}{n},\ \alpha=\beta +\rho.
\ea
\end{itemize}
Definitions \eqref{recINAR1} and \eqref{recBinAR1} use the binomial thinning operator, which is defined by requiring a conditional Bin-distribution, namely $\theta\circ X|X \sim \bin(X,\theta)$ for $\theta\in (0,1)$. Definition \eqref{recIINAR1}, in turn, uses iterated thinning defined as $\rho\circledast_\pi X = \sum_{i=1}^{(\pi\rho)\circ X} Y_i$, where the counting series $(Y_i)$ is \iid\ according to $Y_i\sim 1+\nb(1,\pi)$. The crucial point for the subsequent research is the following:

\smallskip
Models \eqref{recINAR1}--\eqref{recBinAR1} lead to stationary Markov chains with AR$(1)$-like autocorrelation function (ACF) $\rho(h)=\rho^h$ for time lags $h\in\bbn$, where $\rho\in (0,1)$ for all three models. The stationary marginal distributions are $\poi(\mu)$, $\nb(\nu, \frac{\nu}{\nu+\mu})$, and $\bin(n,\mu/n)$, respectively, see \citet{weiss08}. The parameter~$\rho$ allows to control the extent of serial dependence, where the boundary case $\rho\to 0$ leads to \iid\ counts.

\medskip
Independent of the value of~$\rho$, we are always concerned with a Poi-, NB-, or Bin-marginal distribution. Any of these three distributions is uniquely characterized by a corresponding Stein identity, see \citet{sudheesh12}:
\begin{itemize}
	\item $X\sim\poi(\mu)$ if and only if
\ba
\label{SteinChen}
E\big[X\, f(X)\big]\ =\ \mu\, E\big[f(X+1)\big]
\ea
holds for all bounded functions $f:\bbn_0\to\bbr$;
	\item $X\sim\nb(\nu,\frac{\nu}{\nu+\mu})$ if and only if
\ba
\label{SteinNB}
(\nu+\mu)\,E\big[X\,f(X)\big]\ =\ \mu\, E\big[(\nu+X)\,f(X+1)\big]
\ea
holds for all bounded functions $f:\bbn_0\to\bbr$;
	\item $X\sim\bin(n,\mu/n)$ if and only if
\ba
\label{SteinBin}
(n-\mu)\,E\big[X\,f(X)\big] = \mu\,E\big[(n-X)\,f(X+1)\big]
\ea
holds for all bounded functions $f:\bbn_0\to\bbr$.
\end{itemize}
Identity \eqref{SteinChen} is commonly referred to as the Stein--Chen identity \citep[see][]{chen75}. Note that the identities \eqref{SteinChen}--\eqref{SteinBin} constitute non-trivial statements only if~$f$ is not constant on~$\bbn_0$, and if~$f$ is not identical to zero on~$\bbn$. 

\smallskip
In the subsequent Section~\ref{Stein EWMA Charts for Counts}, we shall derive control charts from the identities \eqref{SteinChen}--\eqref{SteinBin}. As these identities have to hold for all bounded functions~$f$ under in-control conditions, we can select any choice of~$f$ for defining the control charts. This degree of freedom can be used to achieve particular sensitivity regarding a specified type of out-of-control scenario.

\section{Stein EWMA Charts for Counts}
\label{Stein EWMA Charts for Counts}
The three identities \eqref{SteinChen}--\eqref{SteinBin} always depend on three types of moment: the mean~$\mu$, the moment $E\big[X\,f(X)\big]$, and a moment involving $f(X+1)$. The idea of the GoF-tests in \citet{aleksandrov22,aleksandrov22b,weissetal23} as well as of the Poisson EWMA charts in \citet{weiss23} was to derive a statistic by solving the identities in a certain way, and to substitute the involved population moments by appropriate types of sample moments. In \citet{weiss23}, two types of constructing an EWMA control chart were analyzed, and it turned out that one of these types is clearly superior, namely the one called ``ABC-EWMA chart''. Building on this experience, we propose the following Stein EWMA charts for sequentially monitoring Poi-, NB-, or Bin-counts, respectively, where $E_0[\cdot]$ expresses that the expectation is computed with respect to the in-control model. For all three Stein EWMA charts, we compute
\ba
\label{AC}
\begin{array}{@{}l}
A_0 = E_0\big[X\,f(X)\big],\qquad A_t\ =\ \lambda\cdot X_t\,f(X_t)\ +\ (1-\lambda)\cdot A_{t-1},
\\[1ex]
C_0 = \mu_0,\qquad C_t\ =\ \lambda\cdot X_t\ +\ (1-\lambda)\cdot C_{t-1},\quad\text{for } t=1,2,\ldots
\end{array}
\ea
Furthermore, we compute for the

\begin{itemize}
	\item $\poi(\mu_0)$ in-control model:
\ba
\label{SEWMApoi}
\begin{array}{@{}l@{\quad}l}
B_0 = E_0\big[f(X+1)\big], & B_t\ =\ \lambda\cdot f(X_t+1)\ +\ (1-\lambda)\cdot B_{t-1},
\\[1ex]
Z_0^{\textup{S}} = 1, & \displaystyle Z_t^{\textup{S}}\ =\ \frac{A_t}{B_t\,C_t},
\qquad\text{for } t=1,2,\ldots;
\end{array}
\ea
	\item $\nb(\nu, \frac{\nu}{\nu+\mu_0})$ in-control model:
\ba
\label{SEWMAnb}
\begin{array}{@{}l@{\quad}l}
\multicolumn{2}{@{}l}{B_0 = E_0\big[(\nu+X)\,f(X+1)\big],}\\[.5ex]
\multicolumn{2}{@{}l}{B_t\ =\ \lambda\cdot (\nu+X_t)\,f(X_t+1)\ +\ (1-\lambda)\cdot B_{t-1},}
\\[1ex]
Z_0^{\textup{S}} = 1, & \displaystyle Z_t^{\textup{S}}\ =\ \frac{(\nu+C_t)\,A_t}{B_t\,C_t},
\qquad\text{for } t=1,2,\ldots;
\end{array}
\ea
	\item $\bin(n,\mu_0/n)$ in-control model:
\ba
\label{SEWMAbin}
\begin{array}{@{}l@{\quad}l}
\multicolumn{2}{@{}l}{B_0 = E_0\big[(n-X)\,f(X+1)\big],}\\[.5ex]
\multicolumn{2}{@{}l}{B_t\ =\ \lambda\cdot (n-X_t)\,f(X_t+1)\ +\ (1-\lambda)\cdot B_{t-1},}
\\[1ex]
Z_0^{\textup{S}} = 1, & \displaystyle Z_t^{\textup{S}}\ =\ \frac{(n-C_t)\,A_t}{B_t\,C_t},
\qquad\text{for } t=1,2,\ldots
\end{array}
\ea
\end{itemize}
We define the respective \emph{Stein EWMA chart} by plotting the statistics $Z_t^{\textup{S}}$ against appropriately chosen $\textup{LCL} < 1 < \textup{UCL}$, where we expect~$Z_t^{\textup{S}}$ to vary closely around~1 under in-control assumptions. 

\medskip
The actual chart design now comprises two steps. First, we have to select the function~$f$ involved in \eqref{AC}--\eqref{SEWMAbin}. Here, one could generally choose any bounded function on~$\bbn_0$, recall \eqref{SteinChen}--\eqref{SteinBin}, except trivial choices like~$f$ being constant on~$\bbn_0$, or~$f\equiv 0$ on~$\bbn$. But not any choice of~$f$ will lead to an appealing chart performance. Instead, $f$ has to be chosen with respect to the anticipated out-of-control scenario, in an analogous way as proposed by \citet{aleksandrov22,aleksandrov22b,weissetal23} and \citet{weiss23}. The idea is to interpret~$f$ as a weight function within the ``A- and B-moments'', which puts unequal weight on the integers in~$\bbn_0$. For a given anticipated out-of-control scenario, one should put most weight on those regions of~$\bbn_0$ where one gets the strongest departures from the in-control model. If we want to uncover zero inflation, for example, we should choose an~$f$ with relatively large weight close to zero counts, whereas for ``general'' overdispersion (\ie if the probability mass function (PMF) is flattened compared to the in-control model), it is advantageous to put more weight on large counts. So in accordance to the recommendations in \citet{weiss23}, the following choices of~$f$ are considered in these two cases:

\begin{itemize}
	\item for uncovering overdispersion relative to the in-control model, we use the linear weights $f(x)=|x-1|$;
	\item for uncovering zero inflation relative to the in-control model, we use the root weights $f(x)=|x-1|^{1/4}$.
\end{itemize}
The case of underdispersion has not been investigated by \citet{weiss23}. Thus, as a starting point, we follow the findings of \citet{weissetal23} on GoF-tests and
\begin{itemize}
	\item for uncovering underdispersion relative to the in-control model, we use the inverse weights $f(x)=1/(x+1)$.
\end{itemize}
The described approach for choosing the weight function is later illustrated in some more detail when discussing Figure~\ref{figPoiGoodpmf}. 
The ARL performance of the aforementioned choices of~$f$ is analyzed in Section~\ref{Simulation-based Performance Analyses} below. As we shall see, the underdispersion scenario is much more demanding than overdispersion or zero inflation. Therefore, further choices for~$f$ shall be considered later in Section~\ref{Underdispersion}. 

\smallskip
After having specified~$f$, the second step of chart design is the choice of the triple $(\lambda, \textup{LCL}, \textup{UCL})$. Assume for the moment that the smoothing parameter~$\lambda$ has already been fixed. Then, LCL and UCL are determined based on ARL considerations. While different ARL concepts exist in the literature \citep{knoth06}, it is common to use the zero-state ARL for evaluating the in-control performance. The most simple solution (the one used here) are symmetric CLs of the form $\textup{LCL} =1-L$ and $\textup{UCL} = 1+L$, where $L$ is chosen such that the desired in-control ARL (ARL$_0$) is met in close approximation (the textbook choice is the target value~$370$). Alternatively, one could define asymmetric CLs by also considering the out-of-control ARL performance, but this is only reasonable if a particular out-of-control scenario has been fixed. For example, if one assumes that the out-of-control models differ from the in-control one solely in terms of the mean~$\mu$, then asymmetric CLs may allow to obtain an unbiased ARL performance with respect to~$\mu$, \ie the ARL as a function of~$\mu$ is maximal in the in-control mean~$\mu_0$ and decreases symmetrically around~$\mu_0$, see Section~4 in \citet{morais20}. But as we shall consider a broad variety of out-of-control scenarios, we restrict to symmetric CLs here, \ie $\mu_0\mp L$ for ordinary EWMA and $1\mp L$ for Stein EWMA charts. 

\smallskip
At this point, let us recall that the above approach for choosing~$f$ assumes that we have specified a \emph{single} relevant out-of-control scenario, such as ``zero inflation''. If the application context allows for \emph{various} out-of-control scenarios (\eg if underdispersion could also be possible), one can run multiple Stein EWMA charts in parallel, each designed for a different type of deterioration. This is later done in Section~\ref{An Illustrative Data Example} when monitoring the emergency counts. There, one Stein EWMA chart is designed to detect overdispersion, another one for zero inflation, and a third one for underdispersion. Then, the observed pattern of alarms enables a kind of targeted diagnosis. For example, if the ``zero inflation''-chart is the first to trigger an alarm, we conclude that we might be confronted with zero inflation (rather than with underdispersion etc.). Certainly, if running many charts in parallel, the risk increases that one of these charts gives a false alarm (``multiple testing''). Generally, the risk of false alarms is controlled by setting a target value ARL$_0$ for the in-control ARL. As described before, in our simulation study, we set ARL$_0\approx 370$ for each single chart for comparability. But if multiple control charts are applied simultaneously, it would also be possible to determine their CLs such that the \emph{joint} in-control ARL meets a target value, \ie the ARL would be defined as the expected time until one of the charts triggers the first alarm.

\smallskip
Another remark refers to the choice of the smoothing parameter~$\lambda$. To ensure a sufficient memory, one often chooses small values for~$\lambda$, such as $\lambda\in\{0.25, 0.10, 0.05\}$. The actual choice of~$\lambda$ might be done based on out-of-control ARLs. For example, we could first compute several candidate triples $(\lambda_i, \textup{LCL}_i, \textup{UCL}_i)$ having the same ARL$_0$. Then, we fix a specified out-of-control scenario and select that triple where the corresponding out-of-control ARL is closest to a given target value. In this way, one might end up with different~$\lambda$ for different Stein EWMA charts. But as we choose $f(x)$ with respect to different out-of-control scenarios, comparable chart designs are most easily obtained by a unique~$\lambda$. Furthermore, all Stein EWMA charts follow the same type of recursive scheme, which makes it plausible to use a unique value for~$\lambda$.  Thus, in the sequel, we follow the practice in \citet{weiss23} and set $\lambda=0.10$ throughout our analyses.

\smallskip
Let us conclude this section with a note on out-of-control ARLs. As mentioned before, there are a couple of competing ARL concepts for evaluating a control chart's out-of-control performance, see \citet{knoth06}, which are typically defined with respect to different positions of the change point. Here, the change point~$\tau$ expresses the time where the process turns out of control, \ie the process is in control (out of control) for $t<\tau$ ($t\geq\tau$). The out-of-control zero-state ARL assumes that the process change happens right at the beginning of process monitoring, \ie the change point equals $\tau=1$. The conditional expected delay CED$(\tau)$, by contrast, assumes a later change point $\tau>1$, where the limit $\tau\to\infty$ leads to the steady-state ARL. While $\tau=1$ would happen for a misspecified in-control model, late change points are often more realistic in practice (but one does not know the true value of~$\tau$ in advance). Fortunately, it turned out in the initial analyses of \citet{weiss23} that the exact position of~$\tau$ does not have a substantial effect on the computed out-of-control ARL values, \ie the Stein EWMA charts showed roughly the same performance for early and late process changes. Since the charts of the present study are defined in complete analogy to those of \citet{weiss23}, we expect the value of the zero-state ARL to be representative also for $\tau>1$ (for ``problematic'' types of control chart, see the discussion in \citet{knoth22}). For this reason, we restrict our subsequent analyses to zero-state ARLs, and these are computed based on simulations. So one simulates the considered process for $R$ times (we use $R=10^4$), applies the specified control chart to it, and determines the time until the first alarm. In this way, one gets $R$ run lengths $l_1,\ldots,l_R$, and the sample mean thereof provides an approximate value of the actual ARL.

\section{Simulation-based Performance Analyses}
\label{Simulation-based Performance Analyses}

\subsection{Overdispersion or Zero Inflation}
\label{Overdispersion or Zero Inflation}
Let us start our performance analyses with the case of overdispersion or zero inflation (compared to the in-control model). Strictly speaking, a zero-inflated distribution also exhibits increased dispersion, but this increase is caused by a single point mass in zero. With overdispersion, by contrast, we refer to a flattened PMF compared to the in-control model's PMF. 
For unbounded counts, we use the NB-distribution for generating overdispersion, and the zero-inflated Poisson (ZIP) distribution for zero inflation. In the bounded-counts case, we use the beta-binomial (BB) distribution for overdispersion and the zero-inflated binomial (ZIB) distribution for zero inflation, see Appendix~A in \citet{weiss18} for details on these distributions. For the sake of a unique representation, we specify any of the aforementioned models by the mean~$\mu$ and by an appropriate dispersion index, namely by $I_{\textup{P}}=\sigma^2/\mu$ for unbounded counts and $I_{\textup{B}}=n\sigma^2/\big(\mu(n-\mu)\big)$ for bounded counts. Note that $I_{\textup{P}}=1$ for the Poi-distribution while $I_{\textup{P}}>1$ for NB and ZIP. Analogously, $I_{\textup{B}}=1$ for the Bin-distribution and $I_{\textup{B}}>1$ for BB and ZIB. The in-control means are set at either $\mu_0=2$ (low counts) or $\mu_0=5$ (medium counts). The data-generating process (DGP) produces either \iid\ counts ($\rho=0$) or AR$(1)$-like counts with $\rho=0.5$, recall \eqref{recINAR1}--\eqref{recBinAR1}. The simulated ARLs rely on $10^4$ replications, the target value for ARL$_0$ is~$370$.

\begin{table}[t!]
\centering
\caption{Simulated ARLs of ordinary EWMA and Stein EWMA chart for Bin-counts with $n=10$. BB and ZIB with $I_{\textup{B}}=5/3$.}
\label{tabARLsOverBin}

\smallskip
\resizebox{\linewidth}{!}{
\begin{tabular}{ll|rr@{}r|r@{}r@{}r|r@{}r@{}r}
\toprule
$\mu_0$ & \multicolumn{1}{r|}{$\mu=$} & \multicolumn{1}{c@{}}{$\mu_0-0.25$} & \multicolumn{1}{c@{}}{$\mu_0$} & \multicolumn{1}{c|}{$\mu_0+0.25$} & \multicolumn{1}{c@{}}{$\mu_0-0.25$} & \multicolumn{1}{c@{}}{$\mu_0$} & \multicolumn{1}{c|}{$\mu_0+0.25$} & \multicolumn{1}{c@{}}{$\mu_0-0.25$} & \multicolumn{1}{c@{}}{$\mu_0$} & \multicolumn{1}{c}{$\mu_0+0.25$} \\
\midrule
\multicolumn{11}{l}{(a)\quad \iid\ counts} \\
\midrule
 &  & \multicolumn{1}{l}{EWMA} &  & \itshape 0.7805 & \multicolumn{1}{l@{}}{EWMA\textsuperscript{S}} & \multicolumn{1}{c@{}}{$|x-1|$} & \itshape 0.534 & \multicolumn{1}{l@{}}{EWMA\textsuperscript{S}} & \multicolumn{1}{c@{}}{$|x-1|^{1/4}$} & \itshape 0.4235 \\
\midrule
2 & ZIB & 69.2 & 87.9 & 51.6 & 22.7 & 26.1 & 29.2 & 16.6 & 19.1 & 21.9 \\
 & Bin & 171.5 & \bfseries 370.2 & 99.3 & 240.2 & \bfseries 369.5 & 550.6 & 191.5 & \bfseries 370.6 & 671.0 \\
 & BB & 71.5 & 90.0 & 51.8 & 25.9 & 29.2 & 33.3 & 30.8 & 40.5 & 55.6 \\
\midrule
 &  & \multicolumn{1}{l}{EWMA} &  & \itshape 0.974 & \multicolumn{1}{l@{}}{EWMA\textsuperscript{S}} & \multicolumn{1}{c@{}}{$|x-1|$} & \itshape 0.2115 & \multicolumn{1}{l@{}}{EWMA\textsuperscript{S}} & \multicolumn{1}{c@{}}{$|x-1|^{1/4}$} & \itshape 0.0511 \\
\midrule
5 & ZIB & 69.2 & 87.9 & 51.6 & 18.9 & 19.9 & 20.6 & 12.9 & 14.0 & 15.5 \\
 & Bin & 162.5 & \bfseries 369.5 & 164.1 & 307.2 & \bfseries 370.1 & 443.4 & 311.5 & \bfseries 369.5 & 417.2 \\
 & BB & 66.7 & 88.2 & 66.3 & 25.1 & 26.9 & 29.0 & 27.4 & 28.9 & 30.3 \\
\midrule
\multicolumn{11}{l}{(b)\quad AR$(1)$ counts with $\rho=0.5$} \\
\midrule
 &  & \multicolumn{1}{l}{EWMA} &  & \itshape 1.191 & \multicolumn{1}{l@{}}{EWMA\textsuperscript{S}} & \multicolumn{1}{c@{}}{$|x-1|$} & \itshape 0.639 & \multicolumn{1}{l@{}}{EWMA\textsuperscript{S}} & \multicolumn{1}{c@{}}{$|x-1|^{1/4}$} & \itshape 0.568 \\
\midrule
2 & ZIB & 77.9 & 73.8 & 55.8 & 22.9 & 24.3 & 25.3 & 23.8 & 26.9 & 30.0 \\
 & Bin & 384.8 & \bfseries 370.1 & 158.0 & 247.1 & \bfseries 369.7 & 554.4 & 211.9 & \bfseries 371.2 & 634.3 \\
 & BB & 105.0 & 105.7 & 77.2 & 33.0 & 39.5 & 47.0 & 44.7 & 58.9 & 77.1 \\
\midrule
 &  & \multicolumn{1}{l}{EWMA} &  & \itshape 1.493 & \multicolumn{1}{l@{}}{EWMA\textsuperscript{S}} & \multicolumn{1}{c@{}}{$|x-1|$} & \itshape 0.225 & \multicolumn{1}{l@{}}{EWMA\textsuperscript{S}} & \multicolumn{1}{c@{}}{$|x-1|^{1/4}$} & \itshape 0.0528 \\
\midrule
5 & ZIB & 30.8 & 29.9 & 28.0 & 7.9 & 7.7 & 7.3 & 7.1 & 6.9 & 6.5 \\
 & Bin & 258.0 & \bfseries 369.1 & 257.3 & 316.7 & \bfseries 370.9 & 424.1 & 293.9 & \bfseries 370.9 & 421.8 \\
 & BB & 86.3 & 96.1 & 85.9 & 35.3 & 37.6 & 39.5 & 37.9 & 39.4 & 40.7 \\
\bottomrule
\\[-2ex]
\multicolumn{2}{l}{\textbf{Notes:}} & \multicolumn{9}{l}{In-control ARL printed in bold font, CL~$L$ shown in italic font. } \\
\multicolumn{2}{l}{} & \multicolumn{9}{l}{``EWMA'' = ordinary EWMA; ``EWMA\textsuperscript{S}'' = Stein EWMA, } \\
\multicolumn{2}{l}{} & \multicolumn{9}{l}{where weight functions $f(x)=|x-1|$ and $|x-1|^{1/4}$. } \\[2ex]
\cmidrule{3-11}
\multicolumn{2}{l}{\textbf{Interpre-}} & &&&  & \multicolumn{1}{c@{}}{\cellcolor[HTML]{C0C0C0}pure} && \multicolumn{3}{c}{\cellcolor[HTML]{C0C0C0}zero inflation} \\
\multicolumn{2}{l}{\textbf{tation:}} & \multicolumn{3}{c|}{\cellcolor[HTML]{C0C0C0}sole mean change} &  & \multicolumn{1}{c@{}}{\cellcolor[HTML]{C0C0C0}distrib.} && && \\
\multicolumn{2}{l}{} & &&&  & \multicolumn{1}{c@{}}{\cellcolor[HTML]{C0C0C0}change} && \multicolumn{3}{c}{\cellcolor[HTML]{C0C0C0}overdispersion} \\
\cmidrule{3-11}
\end{tabular}}
\end{table}

\begin{table}[t!]
\centering
\caption{Simulated ARLs of ordinary EWMA and Stein EWMA chart for NB-counts with $I_{\textup{P}}=5/3$. ZIP with $I_{\textup{P}}=5/3$, and overdispersed NB (``oNB'') with $I_{\textup{P}}=5/2$.}
\label{tabARLsOverNB}

\smallskip
\resizebox{\linewidth}{!}{
\begin{tabular}{ll|rr@{}r|r@{}r@{}r|r@{}r@{}r}
\toprule
$\mu_0$ & \multicolumn{1}{r|}{$\mu=$} & \multicolumn{1}{c@{}}{$\mu_0-0.25$} & \multicolumn{1}{c@{}}{$\mu_0$} & \multicolumn{1}{c|}{$\mu_0+0.25$} & \multicolumn{1}{c@{}}{$\mu_0-0.25$} & \multicolumn{1}{c@{}}{$\mu_0$} & \multicolumn{1}{c|}{$\mu_0+0.25$} & \multicolumn{1}{c@{}}{$\mu_0-0.25$} & \multicolumn{1}{c@{}}{$\mu_0$} & \multicolumn{1}{c}{$\mu_0+0.25$} \\
\midrule
\multicolumn{11}{l}{(a)\quad \iid\ counts} \\
\midrule
 &  & \multicolumn{1}{l}{EWMA} &  & \itshape 1.156 & \multicolumn{1}{l@{}}{EWMA\textsuperscript{S}} & \multicolumn{1}{c@{}}{$|x-1|$} & \itshape 0.349 & \multicolumn{1}{l@{}}{EWMA\textsuperscript{S}} & \multicolumn{1}{c@{}}{$|x-1|^{1/4}$} & \itshape 0.3146 \\
\midrule
2 & ZIP & 506.6 & 462.0 & 149.7 & 139.4 & 257.2 & 481.5 & 54.2 & 81.0 & 124.9 \\
 & NB & 605.8 & \bfseries 370.7 & 133.1 & 172.1 & \bfseries 370.9 & 892.2 & 154.0 & \bfseries 369.9 & 1001.2 \\
 & oNB & 171.9 & 135.1 & 75.6 & 45.2 & 67.2 & 103.8 & 52.2 & 86.9 & 163.2 \\
\midrule
 &  & \multicolumn{1}{l}{EWMA} &  & \itshape 1.805 & \multicolumn{1}{l@{}}{EWMA\textsuperscript{S}} & \multicolumn{1}{c@{}}{$|x-1|$} & \itshape 0.1554 & \multicolumn{1}{l@{}}{EWMA\textsuperscript{S}} & \multicolumn{1}{c@{}}{$|x-1|^{1/4}$} & \itshape 0.0883 \\
\midrule
5 & ZIP & 342.6 & 407.5 & 261.9 & 116.2 & 143.9 & 170.2 & 22.7 & 24.7 & 26.8 \\
 & NB & 444.7 & \bfseries 370.8 & 205.3 & 267.3 & \bfseries 369.8 & 522.7 & 260.0 & \bfseries 370.1 & 537.2 \\
 & oNB & 130.2 & 124.7 & 94.8 & 42.9 & 51.3 & 59.1 & 55.5 & 68.7 & 87.3 \\
\midrule
\multicolumn{11}{l}{(b)\quad AR$(1)$ counts with $\rho=0.5$} \\
\midrule
 &  & \multicolumn{1}{l}{EWMA} &  & \itshape 1.855 & \multicolumn{1}{l@{}}{EWMA\textsuperscript{S}} & \multicolumn{1}{c@{}}{$|x-1|$} & \itshape 0.45 & \multicolumn{1}{l@{}}{EWMA\textsuperscript{S}} & \multicolumn{1}{c@{}}{$|x-1|^{1/4}$} & \itshape 0.4415 \\
\midrule
2 & ZIP & 1287.7 & 505.7 & 238.0 & 108.3 & 167.0 & 266.3 & 90.6 & 139.7 & 219.4 \\
 & NB & 770.9 & \bfseries 369.7 & 200.4 & 187.4 & \bfseries 370.7 & 710.6 & 182.7 & \bfseries 370.7 & 768.8 \\
 & oNB & 288.9 & 178.8 & 115.2 & 61.1 & 93.5 & 141.8 & 75.4 & 123.1 & 210.1 \\
\midrule
 &  & \multicolumn{1}{l}{EWMA} &  & \itshape 2.78 & \multicolumn{1}{l@{}}{EWMA\textsuperscript{S}} & \multicolumn{1}{c@{}}{$|x-1|$} & \itshape 0.177 & \multicolumn{1}{l@{}}{EWMA\textsuperscript{S}} & \multicolumn{1}{c@{}}{$|x-1|^{1/4}$} & \itshape 0.1105 \\
\midrule
5 & ZIP & 502.7 & 408.0 & 284.2 & 147.8 & 178.4 & 214.4 & 93.6 & 110.2 & 131.1 \\
 & NB & 510.8 & \bfseries 369.6 & 242.5 & 278.0 & \bfseries 370.8 & 494.2 & 276.7 & \bfseries 370.2 & 492.8 \\
 & oNB & 183.8 & 156.4 & 124.4 & 60.1 & 71.6 & 83.0 & 78.5 & 97.1 & 122.0 \\
\bottomrule
\\[-2ex]
\multicolumn{2}{l}{\textbf{Notes:}} & \multicolumn{9}{l}{In-control ARL printed in bold font, CL~$L$ shown in italic font. } \\
\multicolumn{2}{l}{} & \multicolumn{9}{l}{``EWMA'' = ordinary EWMA; ``EWMA\textsuperscript{S}'' = Stein EWMA, } \\
\multicolumn{2}{l}{} & \multicolumn{9}{l}{where weight functions $f(x)=|x-1|$ and $|x-1|^{1/4}$. } \\[2ex]
\cmidrule{3-11}
\multicolumn{2}{l}{\textbf{Interpre-}} & &&&  & \multicolumn{1}{c@{}}{\cellcolor[HTML]{C0C0C0}pure} && \multicolumn{3}{c}{\cellcolor[HTML]{C0C0C0}zero inflation} \\
\multicolumn{2}{l}{\textbf{tation:}} & \multicolumn{3}{c|}{\cellcolor[HTML]{C0C0C0}sole mean change} &  & \multicolumn{1}{c@{}}{\cellcolor[HTML]{C0C0C0}distrib.} && && \\
\multicolumn{2}{l}{} & &&&  & \multicolumn{1}{c@{}}{\cellcolor[HTML]{C0C0C0}change} && \multicolumn{3}{c}{\cellcolor[HTML]{C0C0C0}overdispersion} \\
\cmidrule{3-11}
\end{tabular}}
\end{table}

\smallskip
As the case of overdispersion or zero inflation relative to an in-control Poi-distribution was already analyzed by \citet{weiss23}, we restrict our subsequent analyses to in-control NB- and Bin-models, see Tables~\ref{tabARLsOverBin}--\ref{tabARLsOverNB} for the obtained results. The Bin-results are quite close to the Poi-results anyway, which is not surprising as these distributions are related to each other by the Poisson limit theorem. Let us start our discussion with the Bin-case in Table~\ref{tabARLsOverBin}. The design of this table (and of any further table) is as follows. It consists of twelve $3\times 3$-blocks with a negative (positive) mean shift on the left (right) and no mean shift in the center column. In the Bin-rows, the binomial distribution is preserved, whereas we have distributional changes in the ZIB- and BB-rows. Note that at the bottom of each table, there is a scheme for interpreting the ARLs within each block. Part~(a) of each table refers to \iid\ counts, part~(b) to AR$(1)$-like counts. Note that the BB- and ZIB-AR$(1)$ model are constructed as in (3.27) of \citet{weiss18} by replacing the Bin-thinnings in \eqref{recBinAR1} by BB- or ZIB-thinnings, respectively. Finally, the first column of blocks always refers to the ordinary EWMA chart \eqref{ewmarecursion}, whereas the remaining two columns refer to Stein EWMA charts, \ie to \eqref{SEWMAbin} in case of Table~\ref{tabARLsOverBin}. 

\smallskip
Part~(a) of Table~\ref{tabARLsOverBin} just confirms the findings of \citet{weiss23} for the Poi-case. If we are concerned with the textbook situation, \ie if there is a change solely in the mean parameter (the distribution is preserved otherwise), then the ordinary EWMA chart is clearly the best choice. But if there is an additional distributional change, or even only a distributional change while the mean is kept fixed, then the Stein EWMA charts become superior. More precisely, if we are concerned with overdispersion (BB-model), then the weight function $f(x)=|x-1|$ leads to the lowest ARLs, whereas zero inflation is best detected using $f(x)=|x-1|^{1/4}$. It should be noted that we draw the same conclusions from part~(a) of Table~\ref{tabARLsOverNB}, which refers to the NB-case (in-control level of dispersion is $I_{\textup{P}}=5/3$) with Stein EWMA chart \eqref{SEWMAnb}. There, overdispersion (relative to the in-control model) was generated by an NB-distribution with higher dispersion, namely $I_{\textup{P}}=5/2$, whereas we used the ZIP-distribution with $I_{\textup{P}}=5/3$ for causing zero inflation. Thus, we generally conclude that if monitoring \iid\ counts and if being concerned with overdispersion under out-of-control conditions, the Stein EWMA chart with $f(x)=|x-1|$ is the best choice, whereas $f(x)=|x-1|^{1/4}$ performs best under zero inflation. If the type of out-of-control situation is not clear in advance, it is recommended to run all three charts of Tables~\ref{tabARLsOverBin}--\ref{tabARLsOverNB} simultaneously. As these charts show a rather different ARL performance, they can be used for a kind of targeted diagnosis, \ie one can conclude from the observed pattern of alarms on the type of out-of-control situation. For example, if the chart with $f(x)=|x-1|^{1/4}$ signals first, we conclude on zero inflation.

\smallskip
Next, let us look at parts~(b) of Tables~\ref{tabARLsOverBin}--\ref{tabARLsOverNB}, \ie on the case of positively correlated counts. Note that in Table~\ref{tabARLsOverNB}, the overdispersed NB-counts are still generated by the NB-IINAR$(1)$ model \eqref{recIINAR1}, whereas the ZIP-INAR$(1)$ model was used for zero inflation, \ie model \eqref{recINAR1} with ZIP-innovations $(\epsilon_t)$. In most cases in parts~(b), CLs have to be increased and the charts' ARL performances get worse. But it still holds that the Stein EWMA charts are superior under overdispersion or zero inflation. As the only difference, the ARL performances of both Stein EWMA charts are now quite similar under zero inflation in the Bin-case, but not for the NB-case. So altogether, the conclusions done for \iid\ counts still apply, but with generally increased out-of-control ARLs due to the serial dependence.

\begin{figure}[t]
\centering\footnotesize
(a)\hspace{-4ex}\includegraphics[viewport=0 45 295 235, clip=, scale=0.6]{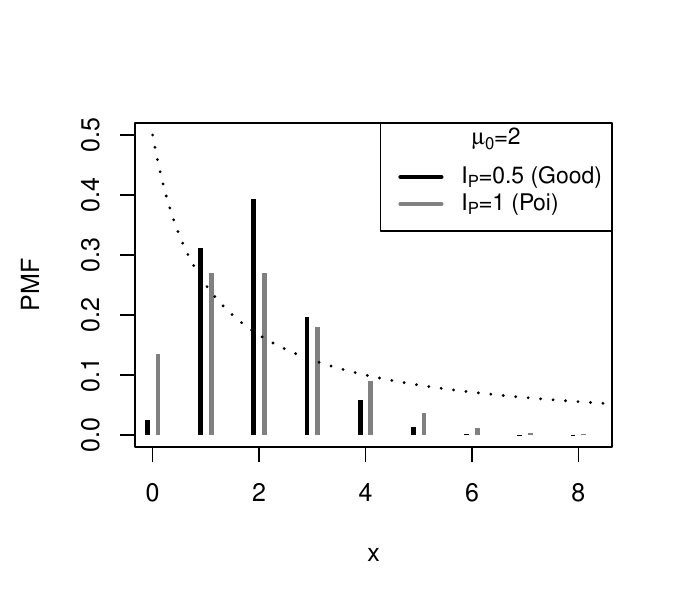}$x$
\quad
(b)\hspace{-4ex}\includegraphics[viewport=0 45 295 235, clip=, scale=0.6]{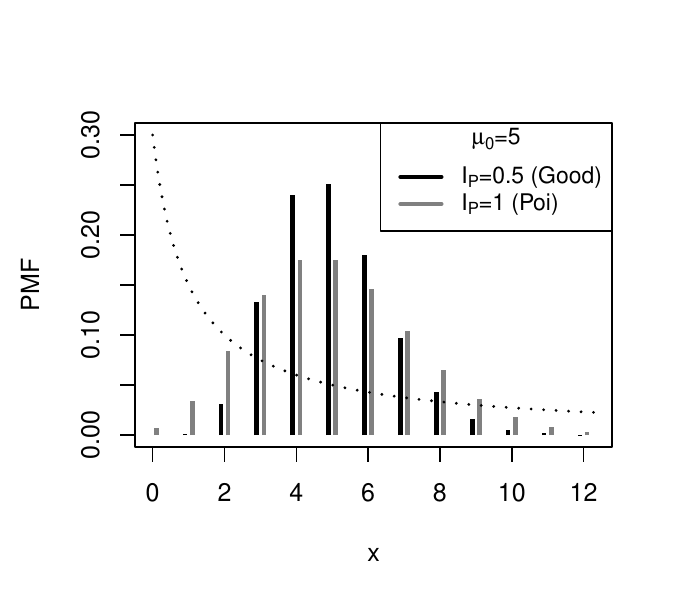}$x$
\caption{PMF plots for Good with $I_{\textup{P}}=0.5$ vs.\ Poi, for (a) $\mu_0=2$ and (b) $\mu_0=5$. Dotted line proportional to $1/(x+1)$.}
\label{figPoiGoodpmf}
\end{figure}

\subsection{Underdispersion}
\label{Underdispersion}
Underdispersion is the opposite phenomenon to overdispersion, \ie the PMF is concentrated more closely around the mean. While there are hardly any models for underdispersion of bounded counts, the underdispersion of unbounded counts received more interest in the literature. For the subsequent simulations, we use the Good distribution \citep[see][Appendix~A]{weiss18} for generating underdispersion relative to the Poi-distribution (see Table~\ref{tabARLsUnderPoi}), whereas for underdispersion relative to an NB-distribution, we use either an NB-distribution with less dispersion or the Poi-distribution (see Table~\ref{tabARLsUnderNB}). It quickly gets clear from Tables~\ref{tabARLsUnderPoi}--\ref{tabARLsUnderNB} that underdispersion is quite demanding for process monitoring. If looking first at the ordinary EWMA chart (first columns of blocks), we recognize severely increasing ARLs in the presence of underdispersion. This is reasonable as fluctuations of the DGP are reduced such that small mean shifts (we have $\pm 0.25$) hardly lead to a violation of the CLs. The second columns in Tables~\ref{tabARLsUnderPoi}--\ref{tabARLsUnderNB} refer to the Stein EWMA charts \eqref{SEWMApoi} and \eqref{SEWMAnb}, where we used $f(x)=1/(x+1)$ as recommended by \citet{weissetal23} in the context of GoF-tests. While this chart indeed performs well for the low mean $\mu=2$, the ARL performance is very bad for $\mu=5$, again demonstrating that process monitoring in the presence of underdispersion is quite demanding. To better understand these difficulties and to find a solution, let us look at Figure~\ref{figPoiGoodpmf}, where we visualize the effect of underdispersion (black) compared to the in-control model (grey) by PMF plots. It can be seen that the strongest deviations are in a small area below~$\mu_0$. For $\mu_0=2$ in (a), the main deviations happen for $x\in\{0,1,2\}$. In fact, $f(x)=1/(x+1)$ (symbolized by the dotted graphs in Figure~\ref{figPoiGoodpmf}) indeed puts most weight to this region, which explains the good ARL performance in this case. For $\mu_0=5$ in (b), however, the main deviations are for $x\in\{2,\ldots,5\}$. As $f(x)=1/(x+1)$ puts relatively low weight into this region, we now get a poor ARL performance.

\begin{table}[t!]
\centering
\caption{Simulated ARLs of ordinary EWMA and Stein EWMA chart for Poi-counts. Good with $I_{\textup{P}}=3/4$ (first row) and $I_{\textup{P}}=1/2$ (second row).}
\label{tabARLsUnderPoi}

\smallskip
\resizebox{\linewidth}{!}{
\begin{tabular}{ll|rr@{}r|r@{}r@{}r|r@{}r@{}r}
\toprule
$\mu_0$ & \multicolumn{1}{r|}{$\mu=$} & \multicolumn{1}{c@{}}{$\mu_0-0.25$} & \multicolumn{1}{c@{}}{$\mu_0$} & \multicolumn{1}{c|}{$\mu_0+0.25$} & \multicolumn{1}{c@{}}{$\mu_0-0.25$} & \multicolumn{1}{c@{}}{$\mu_0$} & \multicolumn{1}{c|}{$\mu_0+0.25$} & \multicolumn{1}{c@{}}{$\mu_0-0.25$} & \multicolumn{1}{c@{}}{$\mu_0$} & \multicolumn{1}{c}{$\mu_0+0.25$} \\
\midrule
\multicolumn{11}{l}{(a)\quad \iid\ counts} \\
\midrule
 &  & \multicolumn{1}{l}{EWMA} &  & \itshape 0.877 & \multicolumn{1}{l@{}}{EWMA\textsuperscript{S}} & \multicolumn{1}{c@{}}{$1/(x+1)$} & \itshape 0.223 & \multicolumn{1}{l@{}}{EWMA\textsuperscript{S}} & \multicolumn{1}{c@{}}{$p_{\textup{P}}(x+2)$} & \itshape 0.608 \\
\midrule
2 & Poi & 252.6 & \bfseries 369.1 & 106.1 & 274.6 & \bfseries 368.9 & 470.8 & 538.9 & \bfseries 370.3 & 271.7 \\
 & Good & 622.4 & 948.9 & 158.5 & 96.9 & 142.3 & 249.5 & 90.7 & 71.4 & 60.5 \\
 & Good & 3611.8 & 6346.9 & 380.6 & 32.3 & 42.2 & 63.0 & 29.1 & 26.2 & 24.3 \\
\midrule
 &  & \multicolumn{1}{l}{EWMA} &  & \itshape 1.388 & \multicolumn{1}{l@{}}{EWMA\textsuperscript{S}} & \multicolumn{1}{c@{}}{$1/(x+1)$} & \itshape 0.1775 & \multicolumn{1}{l@{}}{EWMA\textsuperscript{S}} & \multicolumn{1}{c@{}}{$p_{\textup{P}}(x+2)$} & \itshape 0.293 \\
\midrule
5 & Poi & 309.9 & \bfseries 371.4 & 185.1 & 352.9 & \bfseries 370.5 & 398.1 & 526.1 & \bfseries 368.7 & 268.9 \\
 & Good & 1006.0 & 1015.9 & 327.2 & $>10^4$ & $>10^4$ & $>10^4$ & 228.4 & 149.3 & 106.4 \\
 & Good & 8154.5 & 7882.4 & 1168.9 & $>10^4$ & $>10^4$ & $>10^4$ & 52.5 & 40.1 & 33.2 \\
\midrule
\multicolumn{11}{l}{(b)\quad AR$(1)$ counts with $\rho=0.5$} \\
\midrule
 &  & \multicolumn{1}{l}{EWMA} &  & \itshape 1.351 & \multicolumn{1}{l@{}}{EWMA\textsuperscript{S}} & \multicolumn{1}{c@{}}{$1/(x+1)$} & \itshape 0.2467 & \multicolumn{1}{l@{}}{EWMA\textsuperscript{S}} & \multicolumn{1}{c@{}}{$p_{\textup{P}}(x+2)$} & \itshape 0.7235 \\
\midrule
2 & Poi & 627.2 & \bfseries 371.0 & 162.0 & 274.8 & \bfseries 370.0 & 478.4 & 530.0 & \bfseries 370.5 & 273.2 \\
 & Good & 2177.5 & 814.6 & 261.3 & 153.7 & 242.0 & 439.2 & 143.5 & 108.9 & 89.6 \\
 & Good & $>10^4$ & 2866.0 & 581.6 & 47.6 & 62.4 & 92.3 & 43.7 & 36.7 & 33.2 \\
\midrule
 &  & \multicolumn{1}{l}{EWMA} &  & \itshape 2.123 & \multicolumn{1}{l@{}}{EWMA\textsuperscript{S}} & \multicolumn{1}{c@{}}{$1/(x+1)$} & \itshape 0.1707 & \multicolumn{1}{l@{}}{EWMA\textsuperscript{S}} & \multicolumn{1}{c@{}}{$p_{\textup{P}}(x+2)$} & \itshape 0.345 \\
\midrule
5 & Poi & 432.2 & \bfseries 369.8 & 236.3 & 332.3 & \bfseries 370.1 & 395.3 & 514.5 & \bfseries 370.1 & 280.7 \\
 & Good & 1369.3 & 909.2 & 446.1 & 2652.2 & 4029.5 & 5159.4 & 226.6 & 161.8 & 122.0 \\
 & Good & $>10^4$ & 4264.3 & 1402.0 & 1856.0 & 6087.1 & $>10^4$ & 69.8 & 53.9 & 44.3 \\
\bottomrule
\\[-2ex]
\multicolumn{2}{l}{\textbf{Notes:}} & \multicolumn{9}{l}{In-control ARL printed in bold font, CL~$L$ shown in italic font. } \\
\multicolumn{2}{l}{} & \multicolumn{9}{l}{``EWMA'' = ordinary EWMA; ``EWMA\textsuperscript{S}'' = Stein EWMA, } \\
\multicolumn{2}{l}{} & \multicolumn{9}{l}{where weight functions $f(x)=1/(x+1)$ and $p_{\textup{P}}(x+2)$. } \\[2ex]
\cmidrule{3-11}
\multicolumn{2}{l}{\textbf{Interpre-}} & \multicolumn{3}{c|}{\cellcolor[HTML]{C0C0C0}sole mean change} &  & \multicolumn{1}{c@{}}{\cellcolor[HTML]{C0C0C0}pure} && && \\
\multicolumn{2}{l}{\textbf{tation:}} & &&&  & \multicolumn{1}{c@{}}{\cellcolor[HTML]{C0C0C0}distrib.} && \multicolumn{3}{c}{\cellcolor[HTML]{C0C0C0}increasing} \\
\multicolumn{2}{l}{} & &&&  & \multicolumn{1}{c@{}}{\cellcolor[HTML]{C0C0C0}change} && \multicolumn{3}{c}{\cellcolor[HTML]{C0C0C0}underdispersion} \\
\cmidrule{3-11}
\end{tabular}}
\end{table}

\begin{table}[t!]
\centering
\caption{Simulated ARLs of ordinary EWMA and Stein EWMA chart for NB-counts with $I_{\textup{P}}=5/3$. Underdispersed NB (``uNB'') with $I_{\textup{P}}=4/3$.}
\label{tabARLsUnderNB}

\smallskip
\resizebox{\linewidth}{!}{
\begin{tabular}{ll|rr@{}r|r@{}r@{}r|r@{}r@{}r}
\toprule
$\mu_0$ & \multicolumn{1}{r|}{$\mu=$} & \multicolumn{1}{c@{}}{$\mu_0-0.25$} & \multicolumn{1}{c@{}}{$\mu_0$} & \multicolumn{1}{c|}{$\mu_0+0.25$} & \multicolumn{1}{c@{}}{$\mu_0-0.25$} & \multicolumn{1}{c@{}}{$\mu_0$} & \multicolumn{1}{c|}{$\mu_0+0.25$} & \multicolumn{1}{c@{}}{$\mu_0-0.25$} & \multicolumn{1}{c@{}}{$\mu_0$} & \multicolumn{1}{c}{$\mu_0+0.25$} \\
\midrule
\multicolumn{11}{l}{(a)\quad \iid\ counts} \\
\midrule
 &  & \multicolumn{1}{l}{EWMA} &  & \itshape 1.156 & \multicolumn{1}{l@{}}{EWMA\textsuperscript{S}} & \multicolumn{1}{c@{}}{$1/(x+1)$} & \itshape 0.2215 & \multicolumn{1}{l@{}}{EWMA\textsuperscript{S}} & \multicolumn{1}{c@{}}{$p_{\textup{N}}(x+2)$} & \itshape 0.4163 \\
\midrule
2 & NB & 605.8 & \bfseries 370.7 & 133.1 & 240.3 & \bfseries 371.5 & 590.6 & 367.7 & \bfseries 370.3 & 325.2 \\
 & uNB & 1531.2 & 799.1 & 202.7 & 280.0 & 380.5 & 568.9 & 315.1 & 213.5 & 152.7 \\
 & Poi & 5998.4 & 3237.7 & 429.8 & 106.1 & 128.8 & 179.9 & 93.7 & 70.8 & 57.3 \\
\midrule
 &  & \multicolumn{1}{l}{EWMA} &  & \itshape 1.805 & \multicolumn{1}{l@{}}{EWMA\textsuperscript{S}} & \multicolumn{1}{c@{}}{$1/(x+1)$} & \itshape 0.165 & \multicolumn{1}{l@{}}{EWMA\textsuperscript{S}} & \multicolumn{1}{c@{}}{$p_{\textup{N}}(x+2)$} & \itshape 0.22 \\
\midrule
5 & NB & 444.7 & \bfseries 370.8 & 205.3 & 301.9 & \bfseries 369.1 & 452.9 & 399.3 & \bfseries 369.1 & 328.4 \\
 & uNB & 1072.6 & 840.9 & 372.4 & 1185.9 & 1531.8 & 1921.6 & 455.4 & 313.1 & 226.8 \\
 & Poi & 4236.6 & 3623.7 & 986.3 & 7006.1 & $>10^4$ & $>10^4$ & 132.6 & 96.9 & 76.5 \\
\midrule
\multicolumn{11}{l}{(b)\quad AR$(1)$ counts with $\rho=0.5$} \\
\midrule
 &  & \multicolumn{1}{l}{EWMA} &  & \itshape 1.855 & \multicolumn{1}{l@{}}{EWMA\textsuperscript{S}} & \multicolumn{1}{c@{}}{$1/(x+1)$} & \itshape 0.2412 & \multicolumn{1}{l@{}}{EWMA\textsuperscript{S}} & \multicolumn{1}{c@{}}{$p_{\textup{N}}(x+2)$} & \itshape 0.4626 \\
\midrule
2 & NB & 770.9 & \bfseries 369.7 & 200.4 & 266.3 & \bfseries 369.8 & 514.0 & 440.5 & \bfseries 369.8 & 293.9 \\
 & uNB & 1797.1 & 685.4 & 315.4 & 233.3 & 319.1 & 441.7 & 248.6 & 177.5 & 137.2 \\
 & Poi & $>10^4$ & 2329.1 & 713.9 & 103.7 & 136.9 & 190.5 & 87.6 & 68.6 & 58.2 \\
\midrule
 &  & \multicolumn{1}{l}{EWMA} &  & \itshape 2.78 & \multicolumn{1}{l@{}}{EWMA\textsuperscript{S}} & \multicolumn{1}{c@{}}{$1/(x+1)$} & \itshape 0.1727 & \multicolumn{1}{l@{}}{EWMA\textsuperscript{S}} & \multicolumn{1}{c@{}}{$p_{\textup{N}}(x+2)$} & \itshape 0.247 \\
\midrule
5 & NB & 510.8 & \bfseries 369.6 & 242.5 & 308.7 & \bfseries 370.7 & 442.5 & 414.8 & \bfseries 370.0 & 306.6 \\
 & uNB & 1148.0 & 700.0 & 420.9 & 787.5 & 973.5 & 1222.5 & 329.4 & 241.2 & 191.5 \\
 & Poi & 4918.5 & 2317.8 & 1082.8 & 1689.4 & 2741.1 & 4019.6 & 124.0 & 96.6 & 78.9 \\
\bottomrule
\\[-2ex]
\multicolumn{2}{l}{\textbf{Notes:}} & \multicolumn{9}{l}{In-control ARL printed in bold font, CL~$L$ shown in italic font. } \\
\multicolumn{2}{l}{} & \multicolumn{9}{l}{``EWMA'' = ordinary EWMA; ``EWMA\textsuperscript{S}'' = Stein EWMA, } \\
\multicolumn{2}{l}{} & \multicolumn{9}{l}{where weight functions $f(x)=1/(x+1)$ and $p_{\textup{N}}(x+2)$. } \\[2ex]
\cmidrule{3-11}
\multicolumn{2}{l}{\textbf{Interpre-}} & \multicolumn{3}{c|}{\cellcolor[HTML]{C0C0C0}sole mean change} &  & \multicolumn{1}{c@{}}{\cellcolor[HTML]{C0C0C0}pure} && && \\
\multicolumn{2}{l}{\textbf{tation:}} & &&&  & \multicolumn{1}{c@{}}{\cellcolor[HTML]{C0C0C0}distrib.} && \multicolumn{3}{c}{\cellcolor[HTML]{C0C0C0}increasing} \\
\multicolumn{2}{l}{} & &&&  & \multicolumn{1}{c@{}}{\cellcolor[HTML]{C0C0C0}change} && \multicolumn{3}{c}{\cellcolor[HTML]{C0C0C0}underdispersion} \\
\cmidrule{3-11}
\end{tabular}}
\end{table}

\smallskip
To get a good performance regarding underdispersion, Figure~\ref{figPoiGoodpmf} suggests to put most weight somewhat below the in-control mean, which leads to the following idea. The in-control PMF itself (plotted in grey in Figure~\ref{figPoiGoodpmf}), interpreted as a weight function, puts most weight around the mean of the in-control distribution. Thus, we may just shift this PMF downwards to move most weight below the mean. Inspired by Figure~\ref{figPoiGoodpmf} and some simulation experiments, the idea was developed to define $f(x)$ by simply shifting the in-control PMF by~$2$. More precisely, for the in-control Poi-model of Table~\ref{tabARLsUnderPoi}, we use $p_{\textup{P}}(x+2)$ as the weight function, and $p_{\textup{N}}(x+2)$ for the NB-model of Table~\ref{tabARLsUnderNB}. Here, $p_{\textup{P}}(\cdot)$ and $p_{\textup{N}}(\cdot)$ abbreviate the PMFs of the in-control Poi- and NB-distribution, respectively. The computed chart designs and ARLs are summarized in the respective third column of Tables~\ref{tabARLsUnderPoi}--\ref{tabARLsUnderNB}. We now get notably decreasing ARLs for increasing underdispersion in any scenario, both under \iid\ and AR$(1)$-like counts.

\section{An Illustrative Data Application}
\label{An Illustrative Data Example}
\citet{weisstestik15} analyzed a large set of daily count time series referring to registrations in the emergency department of a children's hospital. More precisely, these emergency counts were determined per 5-min interval from 08:00:00 to 23:59:59 on a day (so length $T=192$), and the full set of time series covers the period from February 13 to August 13, 2009. \citet{weisstestik15} used the sixteen time series from February 13 to 28 to develop the in-control model (Phase-I analysis), namely a Poi-INAR$(1)$ model with in-control mean $\mu_0=2.1$ and dependence parameter $\rho_0=0.78$. Control charts based on this model were then applied to prospective process monitoring (Phase-II application). While most Phase-II series did not contradict the in-control model, \citet{weisstestik15} recognized a few unusual days, two of which shall now serve as illustrative data examples. 

\medskip
Let us start our analyses by designing the control charts based on the in-control model. To make the results consistent with Section~\ref{Simulation-based Performance Analyses}, the CLs of each chart are chosen such that the in-control ARL is close to~370. On the one hand, we consider the c-chart and the ordinary EWMA chart \eqref{ewmarecursion} as competitors. On the other hand, the new Stein EWMA chart \eqref{SEWMApoi} was used together with four weight function: $f(x)=|x-1|$ and $|x-1|^{1/4}$ for indicating possible overdispersion or zero inflation, and $f(x)=1/(x+1)$ and $p_{\textup{P}}(x+2)$ for underdispersion. While the ARLs of the c-chart are computed numerically exactly by using the Markov-chain approach \citep[see][Section~8.2.2]{weiss18}, we again used simulations with $10^4$ replications for the remaining charts. Here, the c-chart is most difficult to design due to discreteness; the best design has $\textup{LCL}=0$ and $\textup{UCL}=6$ (so an only one-sided design, where counts~$X_t>6$ cause an alarm) and ARL$_0 \approx 326.2$. The EWMA chart designs (again with $\lambda=0.1$) are truly two-sided, namely 
\begin{itemize}
	\item EWMA \eqref{ewmarecursion} with $L=1.851$ and ARL$_0\approx 370.3$;
	\item Stein EWMA \eqref{SEWMApoi} with $f(x)=|x-1|$, $L=0.848$, ARL$_0\approx 370.5$;
	\item Stein EWMA \eqref{SEWMApoi} with $f(x)=|x-1|^{1/4}$, $L=0.829$, ARL$_0\approx 370.5$;
	\item Stein EWMA \eqref{SEWMApoi} with $f(x)=1/(x+1)$, $L=0.2994$, ARL$_0\approx 370.5$;
	\item Stein EWMA \eqref{SEWMApoi} with $f(x)=p_{\textup{P}}(x+2)$, $L=0.9594$, ARL$_0\approx 370.2$.
\end{itemize}

\begin{figure}[t!]
\centering\footnotesize
\begin{tabular}{@{}ll@{}}
(a)\quad c-chart: & (b)\quad Ordinary EWMA chart:\\
\includegraphics[viewport=0 45 295 235, clip=, scale=0.6]{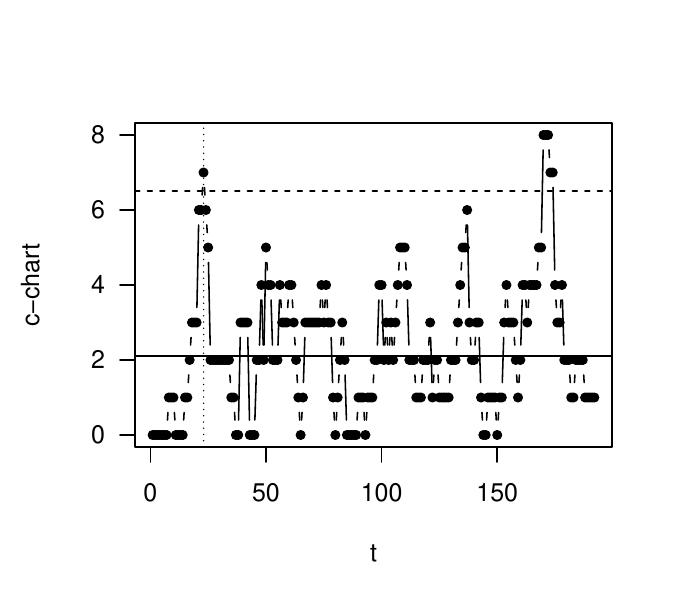}$t$
&
\includegraphics[viewport=0 45 295 235, clip=, scale=0.6]{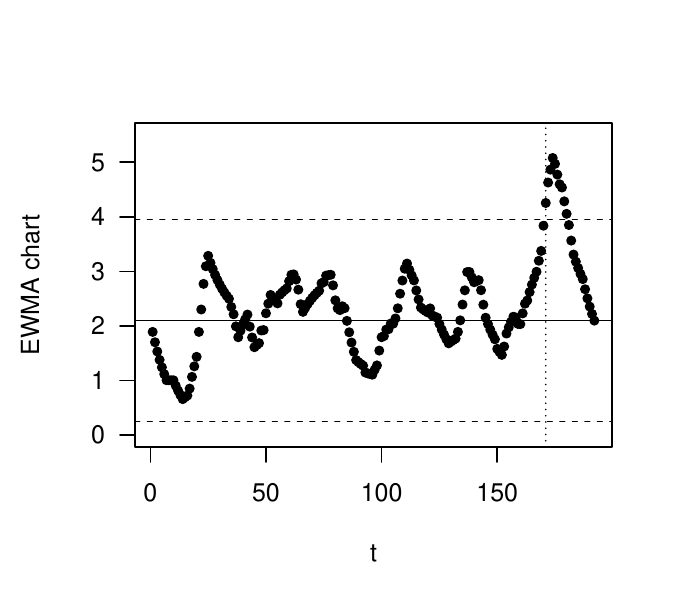}$t$
\\[2ex]
(c)\quad Stein EWMA with $f(x)=|x-1|$: & (d)\quad Stein EWMA with $f(x)=|x-1|^{1/4}$:\\
\includegraphics[viewport=0 45 295 235, clip=, scale=0.6]{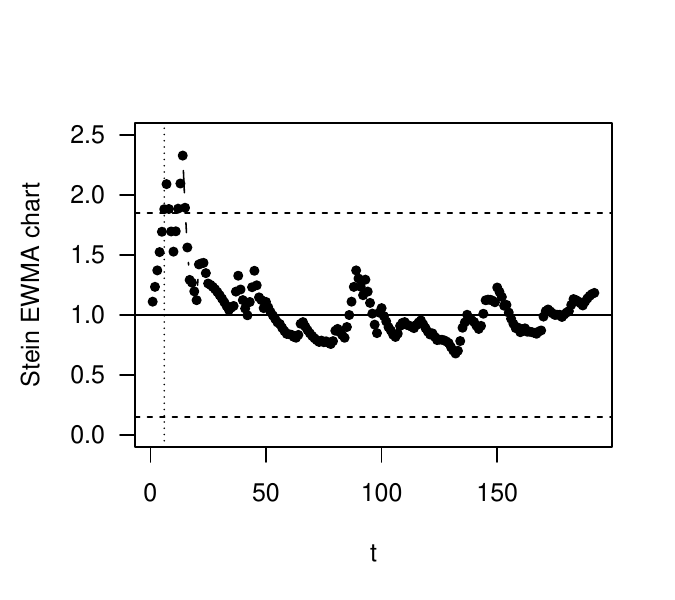}$t$
&
\includegraphics[viewport=0 45 295 235, clip=, scale=0.6]{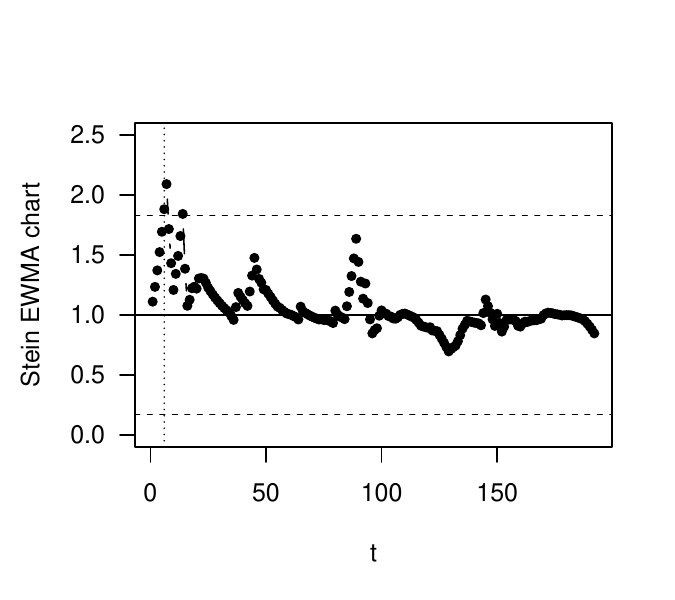}$t$
\\[2ex]
(e)\quad Stein EWMA with $f(x)=1/(x+1)$: & (f)\quad Stein EWMA with $f(x)=p_{\textup{P}}(x+2)$:\\
\includegraphics[viewport=0 45 295 235, clip=, scale=0.6]{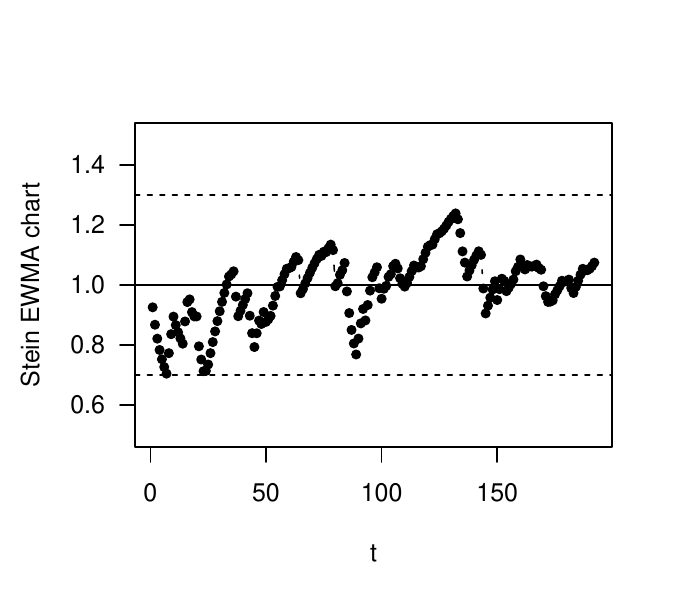}$t$
&
\includegraphics[viewport=0 45 295 235, clip=, scale=0.6]{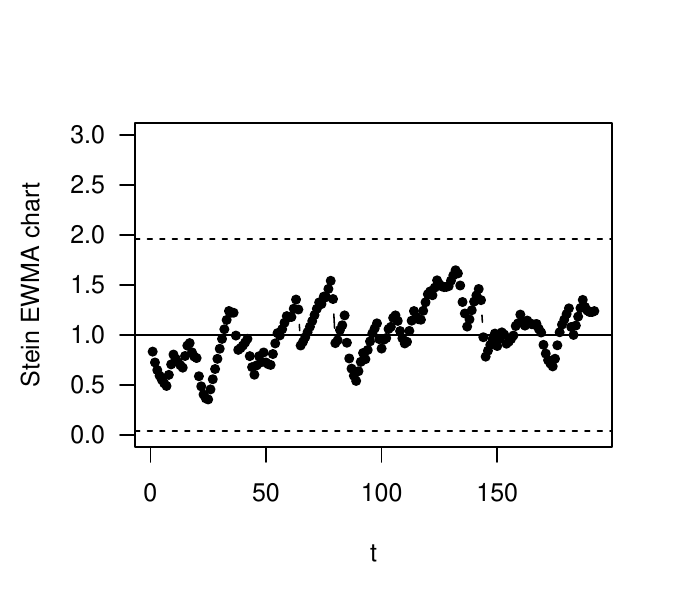}$t$
\end{tabular}
\caption{Emergency counts from March 28, 2009, see Section~\ref{An Illustrative Data Example}: c-chart in (a), ordinary EWMA chart in (b), and different Stein EWMA charts in (c)--(f). CLs as dashed lines, solid center line, and first alarm at dotted line.}
\label{figCharts_2803}
\end{figure}

As a first illustrative example, we apply these control charts to the emergency counts collected on March 28, 2009, see Figure~\ref{figCharts_2803}. The c-chart triggers an alarm at $t=23$, the ordinary EWMA only rather late at $t=171$, both indicating that the in-control model is violated. The situation gets more clear if looking at the four Stein EWMA charts in Figure~\ref{figCharts_2803}. The charts in~(c) and~(d) both trigger a very early alarm at $t=6$, whereas those of~(e) and~(f) do not trigger an alarm at all. This indicates that we are confronted with overdispersion and zero inflation. In fact, comparing the sample properties of the emergency series from March 28, 2009, to the in-control model, we note a slight increase in the mean~$\mu$ (from~2.1 to $\approx 2.323$) and a substantial increase in dispersion~$I_{\textup{P}}$ (from~1 to $\approx 1.312$). Furthermore, the number of zeros equals~27, being larger than~23.5 as expected under the in-control model. So the Stein EWMA charts gave a clear diagnosis of the type of out-of-control situation. In addition, their alarm at $t=6$ was not only much faster than those of c-chart and ordinary EWMA chart, but also than those of the control charts in \citet{weisstestik15} (these trigger at $t=48$ and $t=50$, respectively).

\begin{figure}[t!]
\centering\footnotesize
\begin{tabular}{@{}ll@{}}
(a)\quad c-chart: & (b)\quad Ordinary EWMA chart:\\
\includegraphics[viewport=0 45 295 235, clip=, scale=0.6]{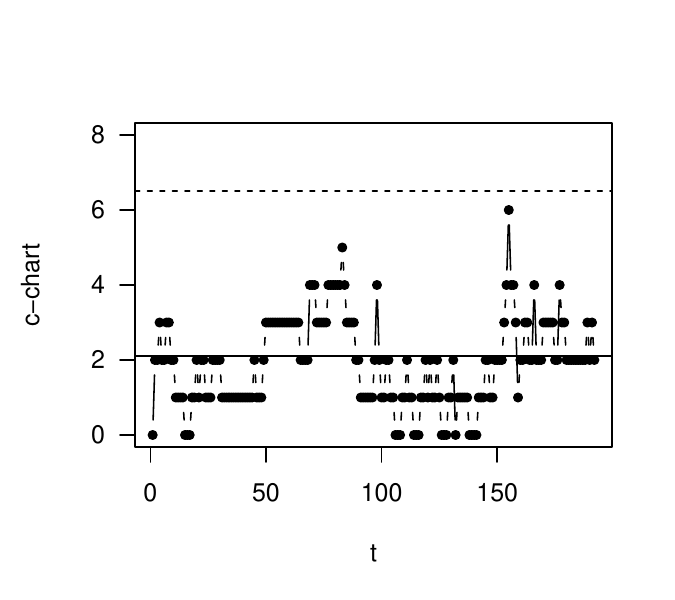}$t$
&
\includegraphics[viewport=0 45 295 235, clip=, scale=0.6]{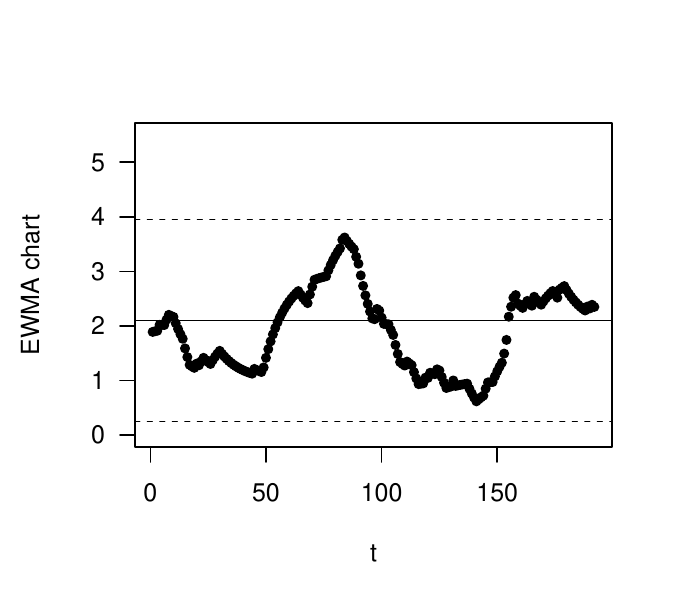}$t$
\\[2ex]
(c)\quad Stein EWMA with $f(x)=|x-1|$: & (d)\quad Stein EWMA with $f(x)=|x-1|^{1/4}$:\\
\includegraphics[viewport=0 45 295 235, clip=, scale=0.6]{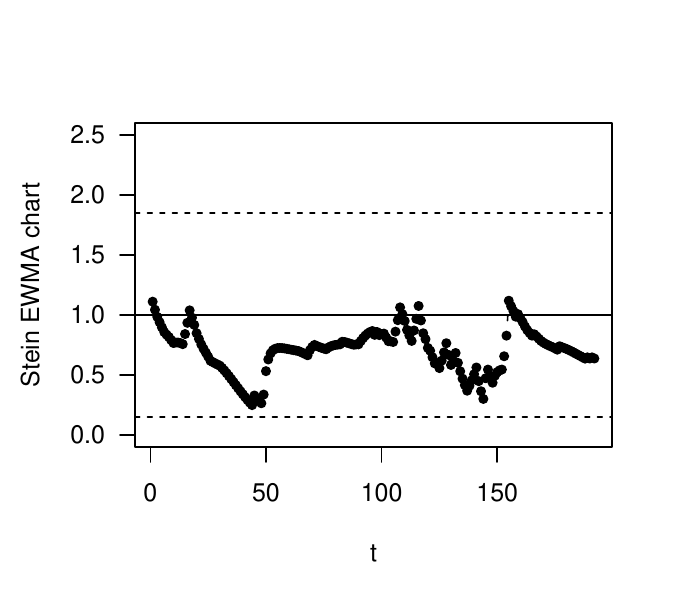}$t$
&
\includegraphics[viewport=0 45 295 235, clip=, scale=0.6]{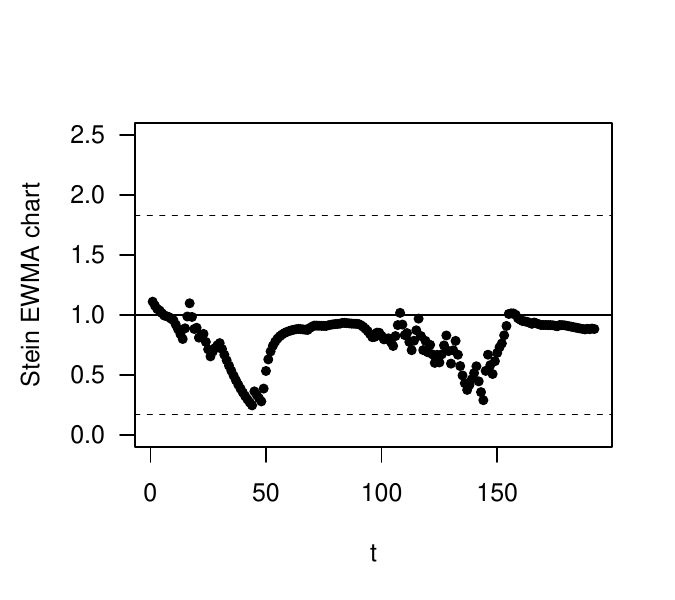}$t$
\\[2ex]
(e)\quad Stein EWMA with $f(x)=1/(x+1)$: & (f)\quad Stein EWMA with $f(x)=p_{\textup{P}}(x+2)$:\\
\includegraphics[viewport=0 45 295 235, clip=, scale=0.6]{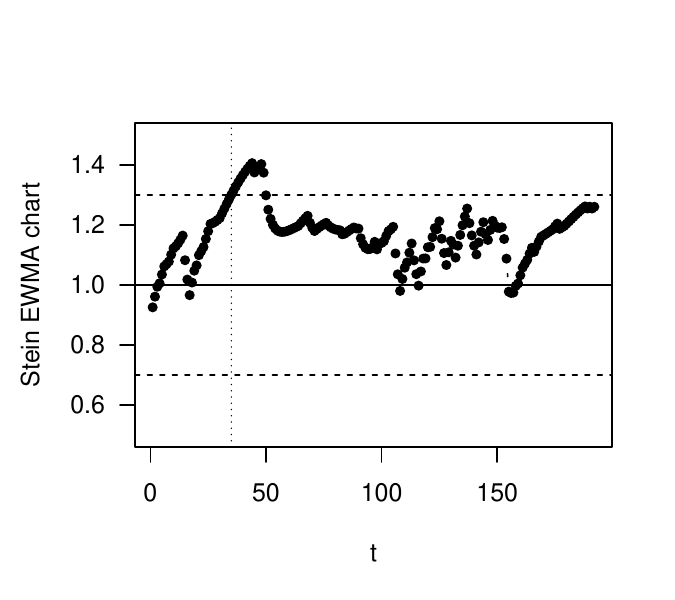}$t$
&
\includegraphics[viewport=0 45 295 235, clip=, scale=0.6]{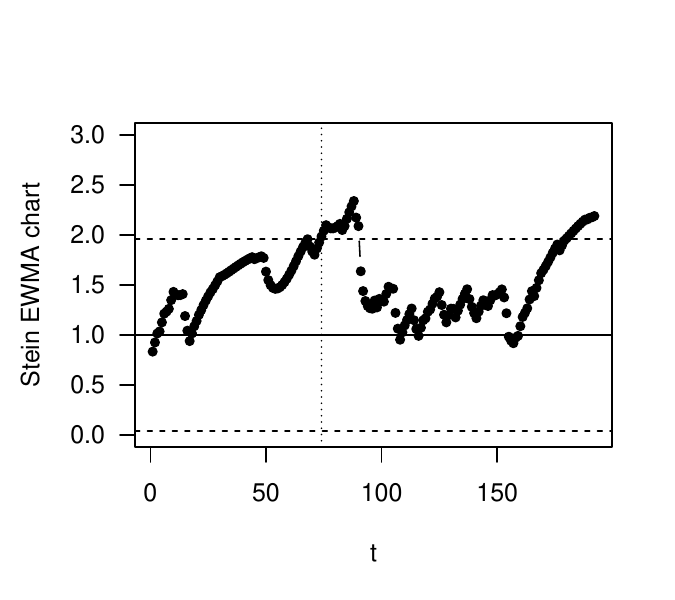}$t$
\end{tabular}
\caption{Emergency counts from July 16, 2009, see Section~\ref{An Illustrative Data Example}: c-chart in (a), ordinary EWMA chart in (b), and different Stein EWMA charts in (c)--(f). CLs as dashed lines, solid center line, and first alarm at dotted line.}
\label{figCharts_1607}
\end{figure}

\smallskip
As the second illustrative example, see Figure~\ref{figCharts_1607}, we consider the emergency counts from July 16, 2009, where no results are reported by \citet{weisstestik15}. This time, neither c-chart nor ordinary EWMA chart trigger an alarm, so the user would conclude that the emergency series is in control. Looking at the Stein EWMA charts, however, the ones in~(e) and~(f) signal at times $t=35$ and $t=74$, respectively, so we seem to be confronted with underdispersion. In fact, the sample value of~$I_{\textup{P}}$ is $\approx 0.710$ being notably smaller than~1, but also the mean has decreased from~2.1 to $\approx 1.911$. As we know from Section~\ref{Underdispersion}, the ordinary EWMA chart is hardly able to detect mean shifts in the presence of underdispersion, so the novel Stein EWMA charts constitute a welcome complement for this scenario.

\section{Conclusions and Future Research}
\label{Conclusions}
For the monitoring of either Poi-, NB-, or Bin-counts, corresponding Stein EWMA charts were proposed, which are constructed by utilizing the respective Stein identity. Their ARL performance was investigated by simulations, both for \iid\ and AR$(1)$-like counts. It turned out that overdispersion is best detected by using the linear weights $f(x)=|x-1|$, and zero inflation by the root weights $f(x)=|x-1|^{1/4}$. Count monitoring in the presence of underdispersion, however, turned out to be quite demanding. While the ordinary EWMA charts performs very poorly in this case, the Stein EWMA chart with inverse weights $f(x)=1/(x+1)$ has appealing out-of-control ARLs for low counts. Even more promising is the weight function obtained by a downward shift of the in-control PMF. These findings were also confirmed by the data example on emergency counts. Nevertheless, it appears that the monitoring of underdispersion requires additional research activity.

\medskip
There are further directions for future research. As cumulative sum (CUSUM) charts often show a better out-of-control performance than EWMA charts, it would be relevant to develop and investigate Stein CUSUM charts for count processes. Here, a residuals-based approach in analogy to \citet{weisstestik15} would be attractive, as this would not only be applicable to Poi-, NB-, or Bin-marginals, but also to in-control models having a Poi-, NB-, or Bin-conditional distribution. Furthermore, Stein EWMA or CUSUM charts would also be interesting for continuously distributed variables data; for this case, Stein identities can be found in \citet{sudheesh09,landsman16}, among others.

\subsubsection*{Acknowledgments}
The author thanks the two referees for their useful comments on an earlier draft of this article. 
This research was funded by the Deutsche Forschungsgemeinschaft (DFG, German Research Foundation) -- Projektnummer 437270842.

\end{document}